\documentclass[twocolumn,twoside,slac_two]{revtex4}
\usepackage{graphicx}
\usepackage{fancyhdr}
\usepackage{graphics}
\usepackage{epstopdf}
\usepackage{textpos}
\begin{document}

\title{\centering Top Quark Production at the Tevatron}
\author{
\centering
\begin{center}
D. Mietlicki \\
(On Behalf of the CDF and D0 Collaborations)
\end{center}}
\affiliation{\centering University of Michigan, Ann Arbor, MI 48109, E-mail: dmietl@fnal.gov}
\begin{abstract}
 The top quark is the most recently discovered of the standard model quarks, and because of 
its very large mass, studies of the top quark and its interactions are important both 
as tests of the standard model and searches for new phenomena.  In this document, 
recent results of analyses of top quark production, via both the electroweak and strong interactions, 
from the CDF and D0 experiments are presented.  The results included here utilize a dataset 
corresponding to up to 6 fb$^{-1}$ of integrated luminosity, slightly more than half of the 
dataset recorded by each experiment before the Tevatron was shutdown in 
September 2011.
\end{abstract}

\maketitle
\thispagestyle{fancy}


\section{INTRODUCTION}
Since its discovery in 1995 at the CDF and DO experiments~\cite{topdiscovery}, the study of the 
production and properties of the top quark has been one of the most active areas of research 
in high energy particle physics.  The top quark has several properties that make it unique 
among quarks, including a very large mass of approximately 172.5~GeV/c$^2$, about 
40 times larger than the next heaviest quark, the $\sim$ 4~GeV/c$^2$ bottom quark.  
Related to this large mass, the top quark also has a very short lifetime.  It decays via 
the electroweak interaction, almost always to $Wb$, with a lifetime that is shorter than 
the time required for hadronization.  As a result, information about 
the top momentum and spin is passed to its decay products, allowing the first opportunity to 
study a ``bare'' quark.

There are two main top quark production mechanisms studied at the Tevatron, where protons and antiprotons 
collided with a center of mass energy of 1.96~TeV.  Individual top quarks can be produced via the 
electroweak interaction, usually in association with additional quark jets.  This 
production mode faces difficulties with large non-top backgrounds and a small cross section, 
and as a result single top production was not observed until 2009~\cite{singletopdiscovery}, 14 years after 
the discovery of the top quark.  The original top quark discovery took place in events where tops are 
produced in pairs by the strong interaction.  The top pair production mode has the advantages 
of a larger cross section and more easily controllable backgrounds, resulting in most analyses of 
top quark properties using top pair events.  For both single top and top pair events, a good 
understanding of the production rate and production mechanisms is essential for studying 
top quark properties, testing the standard model, and looking for new interactions coupling 
to top quarks.

\section{SINGLE TOP QUARK PRODUCTION}

\subsection{Single Top Cross Section}
The electroweak production of single top quarks at the Tevatron is expected to have a cross section 
of approximately 3~pb~\cite{predsingletop}, with the precise value dependent on the mass of the top.  
About one-third of this cross section is due to the exchange of a $W$ 
boson in the s-channel, while the remainder comes from t-channel $W$ exchange.  Single top quarks can 
also be produced in association with a $W$ boson, but the cross section for this process is negligible 
at the Tevatron.

With a small cross section compared to top pair production and a large background from the production of 
$W$ bosons in association with 
jets, single top quarks are difficult to observe experimentally.  Both CDF and D0 use sophisticated 
multivariate techniques such as neural networks and boosted decision trees in order to extract the 
single top production signal.  By combining results using CDF data corresponding to 3.2~fb$^{-1}$ and 
D0 data corresponding to 2.3~fb$^{-1}$, a combined single top cross section of 2.78$^{+0.58}_{-0.47}$~pb is measured, 
which is in good agreement with the theoretical expectation~\cite{tevsingletopxsec}.  Because the single 
top cross section is proportional to the square of the CKM matrix element V$_{tb}$, this cross section 
can be used to extract a 95\% confidence level limit of $|$V$_{tb}| > 0.77$.  D0 also has a more recent 
measurement in 5.4~fb$^{-1}$ which yields a single top cross section of 3.43$^{+0.73}_{-0.74}$~pb~\cite{d0singletopxsec}.

Both CDF and D0 also measure cross sections for s-channel and t-channel single top production separately.  
Such measurements are important tests of the standard model, since many models that predict new 
production mechanisms for single top quarks, such as flavor changing $t-Z-c$ couplings, can alter 
the relative proportion of s-channel and t-channel single top production.  Using a dataset of 3.2~fb$^{-1}$, 
CDF measured cross sections in the two production channels of $\sigma_{s} = 1.8^{+0.7}_{-0.5}$~pb 
and $\sigma_{t} = 0.8^{+0.4}_{-0.4}$~pb~\cite{cdfsingletopxsec}.  A recent D0 analysis with 5.4~fb$^{-1}$ 
found $\sigma_{s} = 0.98\pm0.63$~pb and $\sigma_{t} = 2.90\pm0.59$~pb~\cite{d0tchannelxsec}.  This D0 result provided 
the first observation of single top production via the t-channel exchange of a $W$ boson.

\subsection{Exotic Single Top Production}
\begin{figure}
\includegraphics[width=65mm]{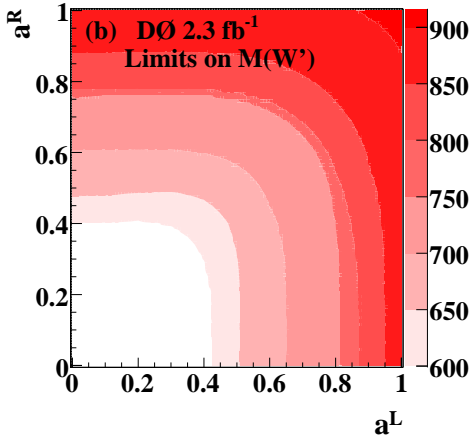}
\caption{$W^{\prime}$ mass limits from D0.}
\label{wprimefig}
\end{figure}

Many proposed extensions to the standard model result in new mechanisms for producing top quarks.  In particular, 
one new mechanism that is common to many new models is the production of a top quark and a bottom quark from the 
decay of a heavier version of the $W$ boson, called a $W^{\prime}$.  Searches for $W^{\prime}$ production can be 
performed by looking for resonant production of a top quark with a $b$.  D0 has recently performed a search for 
$W^{\prime}$ bosons decaying to $tb$, allowing for arbitrary combinations of left-handed and right-handed $W^{\prime}$ 
couplings to fermions~\cite{d0wprime}.  This analysis improved the 95\% confidence level limit 
on the mass of a $W^{\prime}$ with purely right-handed couplings to M$_{W^{\prime}} > 885$~GeV/c$^2$, an increase of 
85~GeV/c$^2$ over the previous limit set by CDF~\cite{cdfwprime}.  Mass limits from D0 for arbitrary left-handed and right-handed 
couplings are shown in Figure~\ref{wprimefig}.

\section{TOP QUARK PAIR PRODUCTION}

\subsection{Cross Section Measurements}
Top pair production via the strong interaction has a predicted cross section of approximately 7.5~pb~\cite{predttbar} at the 
Tevatron.  This larger cross section compared to single top production, along with the fact 
that the signal is more easily distinguished from the backgrounds because of the presence of two top quarks rather than 
one, makes $t\bar{t}$ production an ideal place to study top properties.  However, first the production rate 
of top pairs must be well understood.

Because tops decay predominantly to $Wb$, the $t\bar{t}$ final states are categorized by the decay of the two 
$W$ bosons in the event.  The dilepton decay channel, with both $W$'s decaying to leptons\footnote{Throughout, 
``lepton'' refers only to electrons or muons.  Tau leptons are more difficult to identify experimentally and 
are generally treated separately.}, provides a very clean channel with little background, but the branching 
ratio for this channel is also small and as a result the total number of candidate events is smaller than in other channels.  
The lepton+jets channel, with one $W$ decaying leptonically and the other decaying to a pair of quarks, provides a 
good combination of both relatively small background contamination and a reasonably large branching ratio.  
Because events in this channel are fully constrained kinematically, so that the top quarks can be completely 
reconstructed, this channel is often used in top quark property measurements.  The all-hadronic channel, 
where both $W$ bosons decay to quark pairs, has the largest branching ratio, but also faces extremely large 
backgrounds from QCD multijet production.

The most precise measurements of the top quark pair production cross section at the Tevatron use the 
lepton+jets channel.  Top decays almost always include a $b$ quark, and since hadrons containing $b$'s have a long lifetime, 
jets originating from $b$ quarks can be identified by the presence of displaced secondary vertices from 
long-lived particles.  Cross section measurements in the lepton+jets channel can be made both with and without 
requiring the presence of such $b$-tagged jets.  In the case where no tagged jets are present, kinematic features are used 
to separate the $t\bar{t}$ signal from the backgrounds, while for the case where a $b$-tag is required, 
backgrounds are small enough that a simple counting experiment can be used to measure a cross section.  
CDF has performed measurements using both methods with 4.3~fb$^{-1}$ and then combined the results, yielding a cross section 
of $7.70\pm0.52$~pb~\cite{cdflepjets}.  A recent D0 analysis in 5.4~fb$^{-1}$ combined aspects of both the $b$-tagged and 
non-$b$-tagged approach to find a cross section of 7.78$^{+0.77}_{-0.64}$~pb~\cite{d0lepjets}.

The dilepton channel has a limited sample size due to its small branching ratio, but because of its low background 
contamination, it provides a sample that is almost exclusively made up of top quark pairs.  D0 performed a measurement in this 
channel with 5.1~fb$^{-1}$ that also takes advantage of $b$-tagging information in order to further separate signal from background.  
A neural network was used to determine the probability that the jets in each event originated from $b$ quarks, and then 
a likelihood fit to the resulting discriminant was performed to extract the $t\bar{t}$ cross section.  This procedure 
yielded a cross section of 7.36$^{+0.90}_{-0.79}$~pb~\cite{d0dil}, with the precision beginning to approach that of the 
lepton+jets channel.  CDF also measured the cross section in the dilepton channel with a requirement that one of the jets 
be tagged as coming from a $b$, yielding $7.25\pm0.92$~pb~\cite{cdfdil} in 5.1~fb$^{-1}$.

Top events where one of the $W$ bosons decays to a tau lepton are generally difficult to identify.  If the tau decays to an 
electron or muon, the event can be found in the dilepton or lepton+jets samples, but taus can also decay hadronically, in 
which case it is difficult to distinguish them from quark jets.  In such cases, the tau can be identified as a narrow jet 
containing a small number of charged tracks, but there is a very large background from QCD multijet production.  
Sophisticated multivariate techniques are needed to extract the signal from the background.  A recent CDF result in this 
channel used 2.2~fb$^{-1}$ to find a cross section of $8.8\pm4.3$~pb~\cite{cdftaujet}, which is in good agreement with the 
theoretical prediction and with the results measured in other decay channels.  D0 performed a measurement in the same channel 
using 1~fb$^{-1}$, finding a cross section of 6.9$^{+1.5}_{-1.4}$~pb~\cite{d0taujet}.

Because the CDF and D0 results agree well across different decay channels, the cross section measurements from the various 
channels can be combined in order to improve precision.  Both CDF and D0 have produced combined top pair cross section 
measurements\footnote{The combinations include some results not listed here, and some of the most recent measurements discussed here are not 
included in the combinations.}.  CDF measured a combined cross section of $7.50\pm0.48$~pb~\cite{cdfxsec}.  D0 found a cross 
section of 7.56$^{+0.63}_{-0.56}$~pb~\cite{d0dil} when combining measurements in the dilepton and lepton+jets channels.  
Both experiments measure cross sections that are in good agreement with each 
other and with the standard model prediction.

\subsection{Other Aspects of Top Pair Production}
Beyond inclusive cross section measurements, there are many other aspects of $t\bar{t}$ production that are studied at the 
Tevatron.  Some of these analyses are meant to test specific predictions of the standard model, for example by comparing 
differential cross sections to theoretical expectations.  Others are explicit searches for new particles and interactions 
that can produce top pairs.

\subsubsection{Cross Section vs. Top $p_T$}
The D0 experiment has studied the differential cross section for top pair production as a function of the transverse 
momentum, $p_T$, of the top quark~\cite{d0toppt}.  This variable is of particular interest because in many models containing 
new particles that interact with tops, these new particles are very massive.  As a result, the top quarks produced by these 
interactions would have large $p_T$.  D0 measured the differential cross section in the lepton+jets channel, unfolding to 
correct for acceptance and detector affects, and then compared the result to various calculations of the standard model 
prediction, finding good agreement.

\subsubsection{Boosted Top Quarks}
A recent analysis at CDF searched for highly boosted top quarks with $p_T$ above 400~GeV/c~\cite{cdftopboost}.  Although the 
standard model cross section for such highly boosted tops is small at the Tevatron (approximately 5 fb), this search is 
important both because new interactions could produce very energetic top pairs and also for comparison with LHC data, 
where these boosted tops will be much more common.  At such high energies, the top quark decay products are collimated into a 
single jet, and the analysis relied on the ability to distinguish these energetic top jets from energetic light quark jets.  
This analysis placed a 95\% confidence level upper limit of 38~fb on the cross section for the production of top pairs with at 
least one top having $p_T$ above 400~GeV/c.

\subsubsection{Resonance Searches}
Many models for physics beyond the standard model predict that top quark pairs can be produced via a resonance of some new 
heavy particle that decays to $t\bar{t}$.  For example, new physics models that contain additional gauge symmetries often predict 
a new heavy version of the $Z$ boson, called $Z^{\prime}$, that decays to top pairs.  As long as such resonances are not too 
wide, they can appear as bumps in the $t\bar{t}$ invariant mass distribution at the mass of the new particle.  

Searches have been performed at both CDF and D0 for resonant $t\bar{t}$ production by searching for bumps in the top pair mass 
distribution.  At D0, using 3.6 fb$^{-1}$, a 95\% confidence level limit on the mass of a new $Z^{\prime}$ of M$_{Z^{\prime}} > 820$~GeV/c$^2$ 
was set~\cite{d0resonance}.  A more recent analysis at CDF using 4.8 fb$^{-1}$ set a limit of M$_{Z^{\prime}} > 900$~GeV/c$^2$~\cite{cdfresonance}.

\subsubsection{New Particles Decaying to Top}
Top quark pairs can also be created through the pair production of some heavy particle that then decays to top.  One particular search 
channel where this type of interaction could be observed is the production of top pairs along with additional 
missing transverse energy due to an invisible particle.  This final state could be created, for example, by the pair production 
of exotic fourth generation quarks $T^{\prime}$, with each $T^{\prime}$ then decaying to a top quark and a dark matter particle.

\begin{figure}
\includegraphics[width=65mm]{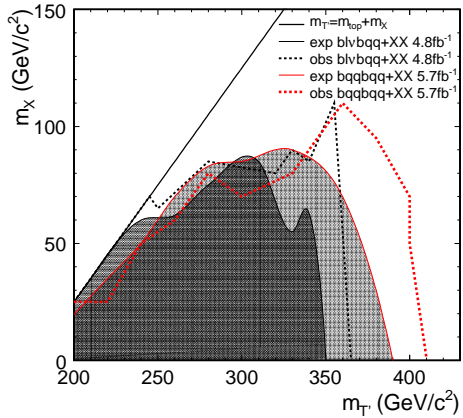}
\caption{Limits on $T^{\prime}$ and dark matter particle masses from CDF.}
\label{ttmet}
\end{figure}

CDF performed two searches for this process, one in the lepton+jets decay channel and one in the all-hadronic channel~\cite{cdftopmet}.  
Although the searches were optimized for this particular process, they were also sensitive to other processes with the same final state - 
for example, supersymmetric stop pair production, with the stop quarks decaying to tops and neutralinos.  After selecting top quark pair 
events, these analyses looked for evidence of large additional missing transverse energy.  No such evidence was found, and limits were placed 
on the masses of the $T^{\prime}$ and the dark matter particle, as shown in Figure~\ref{ttmet}.

\subsubsection{Searches for Heavy Top Quarks}
If a fourth generation of standard-model-like quarks exists, then it is possible that the fourth generation top-like quark, $t^{\prime}$,
could decay to a final state similar to top, $t^{\prime} \rightarrow Wq$ where $q$ is a down-type quark ($d$, $s$, or $b$).  In this case, 
pair-produced $t^{\prime}$ would have the same final states as standard model $t\bar{t}$ production and such events could be found in the 
$t\bar{t}$ signal sample.  Both CDF and D0 performed analyses which search the $t\bar{t}$ sample in the lepton+jets final state for such events 
by looking at both the reconstructed top quark mass and total transverse energy in each event.

At D0, using 5.3~fb$^{-1}$, a 95\% confidence level limit on the mass of the $t^{\prime}$ was set, limiting M$_{t^{\prime}} > 285$~GeV/c$^2$~\cite{d0tprime}.  
This is smaller than the expected limit of 320~GeV/c$^2$ due to a small excess in events where the lepton 
was a muon.  CDF used 5.6~fb$^{-1}$ to set a limit of M$_{t^{\prime}} > 340$~GeV/c$^2$~\cite{cdftprime}.  In the CDF analysis, when the 
additional assumption was made that the down-type quark $q$ is always specifically a $b$ quark, $b$-tagging could be utilized and the 
limit was strengthened to M$_{t^{\prime}} > 358$~GeV/c$^2$.

\subsubsection{Forward-Backward Asymmetry}
One aspect of $t\bar{t}$ production that has been of particular interest is the 
forward-backward asymmetry.  This analysis is particularly suited to a $p\bar{p}$ collider like the Tevatron, rather than 
the LHC which has a symmetric $pp$ initial state.  The standard model predicts that in $t\bar{t}$ production, there will 
be a slight excess in the number of top quarks that follow the proton direction compared to antitops, resulting in an asymmetry 
of approximately 5\%~\cite{predafb}.  Many extensions to the standard model predict an enhancement of the asymmetry.  Both CDF and D0 measure 
this asymmetry, with corrections applied to account for acceptance and detector effects.  CDF performed measurements in both 
the dilepton and lepton+jets decay channels, which when combined yield an asymmetry of $20\pm7$\%~\cite{cdfafb}.  A recent 
D0 measurement found an asymmetry of $19.6\pm6.5$\%~\cite{d0afb}.  The asymmetries from both experiments are  approximately 
three standard deviations away from zero and up to two standard deviations away from theoretical calculations of the standard model asymmetry, 
leading to much interest in the possibility that this asymmetry could be caused by a new top pair production 
mechanism, for example~\cite{afbtheory}.

\section{CONCLUSION}
Studies of the interactions of the top quark remain an important part of the research program at the Tevatron.  Although the 
data-taking run has now ended, the analyses presented here only use up to a little more than half of the recorded data.  
Both the CDF and D0 experiments have much data yet to be analyzed, 
and analyses focused on top production play an integral role in testing the standard model and understanding the 
sample composition for detailed studies of top quark properties.  Precise cross section measurements will continue to 
test the predictions of QCD and establish the top quark backgrounds in searches for new physics.  Analyses of production 
properties that are more difficult to study at the LHC, such as the top pair forward-backward asymmetry, also continue 
to move forward at the Tevatron, furthering our understanding of the most recently discovered quark.

\bigskip 
\begin{acknowledgments}
I would like to thank the members of the CDF and DO collaborations for
their dedication to the study of top quark production, especially the
top group conveners and the authors of the analyses included here.  I 
also want to thank the organizers of the 2011 Physics in Collision 
symposium and everyone who contributed to the conference.
\end{acknowledgments}

\bigskip 
\bibliography{basename of .bib file}

\end{document}